\documentclass[twocolumn,showpacs,preprintnumbers,amsmath,amssymb]{revtex4}
\usepackage{graphicx}% Include figure files
\usepackage{dcolumn}% Align table columns on decimal point
\usepackage{bm}% bold math

\begin{document}
\preprint{APS/123-QED}
   %Fixing abstract in twocolumn mode
   
 %Fixing abstract in twocolumn mode

%\twocolumn[\hsize\textwidth\columnwidth\hsize\csname @twocolumnfalse\endcsname]
\title{Breakdown of the Onsager Principle as a Sign of Aging}

\author{Paolo Allegrini$^{1}$,  Gerardo Aquino$^{2}$, Paolo Grigolini$^{2,3,4}$ , Luigi Palatella$^{3}$, Angelo Rosa$^{5}$\\} 

% \author{Second Author}%
% \email{Second.Author@institution.edu}
\affiliation{$^{1}$ Istituto di Linguistica Computazionale del Consiglio Nazionale delle
Ricerche, Area della Ricerca di Pisa. Via Moruzzi 1,
56124, Pisa, Italy\\
$^{2}$Center for Nonlinear Science, University of North Texas,
   P.O. Box 311427, Denton, Texas 76203-1427\\
$^{3}$Dipartimento di Fisica dell'Universit\`a di Pisa and
INFM, via Buonarroti 2, 56127 Pisa,\\
$^{4}$Istituto di Biofisica
Consiglio Nazionale delle Ricerche
Area della Ricerca di Pisa. Via G. Moruzzi 1
56124 Pisa, Italy\\
$^{5}$  International School For Advanced Studies and INFM
 Via Beirut 2-4, 34014   Trieste,  Italy}
\date{\today}% It is always \today, today,
             %  but any date may be explicitly specified\date{\today}% 

\begin{abstract}

We discuss the problem of the equivalence between Continuous Time Random Walk (CTRW) and
Generalized Master Equation (GME). The walker, making instantaneous jumps from one site of
the lattice to another , resides in each site for extended times. The sojourn times have a distribution
$\psi(t)$ that is assumed to be an inverse power law with the power index $\mu $. We assume that
the Onsager principle is fulfilled, and we use this assumption to establish a complete equivalence
between GME and the Montroll-Weiss CTRW. We prove that this equivalence is confined to the
case when $\psi(t)$ is an exponential. We argue that  is so because the Montroll-Weiss CTRW,
as recently proved by Barkai [E. Barkai, Phys. Rev. Lett. 90, 104101 (2003)], is non-stationary,
thereby implying aging, while the Onsager principle, is valid only in the case of fully aged systems.
The case of a Poissonian distribution of sojourn times is the only one with no aging associated to it,
and consequently with no need to establish special initial conditions to fulfill the Onsager principle.
We consider the case of a dichotomous fluctuation, and we prove that the Onsager principle is
fulfilled for any form of regression to equilibrium provided that the stationary condition holds true.
We set the  stationary condition on both the CTRW and the
GME, thereby creating a condition of total equivalence, regardless the nature of the waiting time
distribution. As a consequence of this procedure we create a GME that it is a \emph{bona fide}
master equation, in spite of being non-Markovian. We note that the memory kernel of the GME
affords information on the interaction between system of interest and its bath. The Poisson case
yields a bath with infinitely fast fluctuations. We argue that departing from the Poisson form has
the effect of creating a condition of infinite memory and that these results might be useful to shed light into the problem of how to unravel non-Markovian master equations.
\end{abstract}
%Valid PACS appear here% PACS, the Physics and Astronomy
                             % Classification Scheme.
%\keywords{Suggested keywords}%Use showkeys class option if keyword
                              %display desired
\maketitle

\section{introduction}
The Onsager principle \cite{onsager} is one of the basic tenets of statistical mechanics insofar as it establishes a connection between a property of equilibrium, the correlation function of a given variable $A$, and the regression to equilibrium of a macroscopic signal. For this reason, we judge the Onsager principle to be  a fundamental step for the connection between dynamics and thermodynamics. It is important to stress, as clearly stated by Onsager himself \cite{onsager}, that this principle holds true for aged systems, namely systems in contact with heat reservoirs that are supposed to be in a condition of thermal equilibrium. If the regression to equilibrium is very fast, it is not so important to ensure the equilibrium condition of the bath, at the moment when we begin measuring the regression to equilibrium of the system of interest. In fact, in the specific case where the bath is responsible for fluctuations that can be assumed to be white, the regression to equilibrium of the bath is instantaneous.
The Onsager principle refers to a variable of interest whose dynamics are made stochastic by the interaction with a bath. Thus, when we discuss the process of regression to equilibrium, we have to specify if we are referring to the system of interest or to its bath. As earlier said, if we adopt the white noise approximation to describe the fluctuations that are responsible for the erratic motion of the variable of interest, then the regression to equilibrium of the reservoir is virtually instantaneous and we can easily fulfill the condition for the validity of the Onsager principle. The variable of interest is characterized by a stationary correlation function $\Phi_{A}(t_{1}, t_{2})  = \langle A(t_{1}) A(t_{2})\rangle $, which only depends on $|t_{1}-t_{2}|$. The stochastic behavior of the variable of interest is caused by the interaction between system of interest and bath, and this kind of process is often studied by means of the master equation method. A popular method to derive the master equation is the projection method by Zwanzig \cite{zwanzig}. However, this method is easy and convenient to use, when special initial conditions can be adopted, with the total distribution expressed as the product of the relevant, or reduced distribution,  and of the bath distribution \cite{grigo}.
These special initial conditions have the beneficial effect of annihilating the inhomogeneous term
that makes the reduced equation of motion explicitly dependent on the initial condition \cite{grigo}. If no initial condition of this kind is adopted, we have to consider also the inhomogeneous term and we have to wait a time comparable to the bath relaxation time for this term to disappear. This suggests that a non-Markov master equation with the same structure as the Zwanzig master equation, with no inhomogeneous term, does not necessarily refer to a stationary condition. It might refer to an initial condition with system of interest and bath uncorrelated one from the other, a non-stationary state. As we shall see, to create a master equation compatible with the Onsager principle, in the case when the bath is not infinitely fast, we must establish a delicate condition of entanglement between system and bath. This is not a trivial problem, since the departure from the Poisson condition generates an infinitely extended memory, and the system-bathentanglement is the result of a rearrangement process with an infinite time scale.

 In the last few years there has been a growing interest for the formalism of Continuous Time Random Walk (CTRW) \cite{montrollweiss}, because of its close connection with the adoption of fractional operators that, in turn, are revealing a powerful tool to describe cooperative processes in condensed matter \cite{bruce}. On the other hand, the connection between the master equation formalism and the CTRW walk has been discussed over the years, beginning with the pioneering work of Ref.\ \cite{katja}. According to the CTRW the walker travelling through an one dimensional path makes jumps from one site to the others after spending a given amount of time in the departure site. The distribution of sojourn times is assumed to be the same for all the sites, and is denoted by the function $\psi(t)$. The pioneering work of Ref.\ \cite{katja} proves that the Markovian master equation is compatible with  the CTRW, if  $\psi(t)$ is a Poissonian distribution. \\ 

        The results of the pioneering work of Ref.\ \cite{katja} raise the important issue of the connection between CTRW, with non-Poissonian $\psi(t)$, and a non-Markovian master equation.  Apparently, this important problem is solved by adopting the non-Markovian master equation proposed by the authors of Ref.\ \cite{kenkre}. We consider this theoretical tool to be very important, and we denote with the term Generalized Master Equation (GME) all transport equations with the same structure. Recently the GME has been discussed by Metzler \cite{metzler} who argued that this equation unifies fractional calculus and CTRW. We want to remark, however, that here we focus our attention on how to make the GME stationary, and thus compatible with the Onsager principle. In fact, as earlier remarked, the absence of an inhomogeneous term does not guarantee the stationary nature of the transport equation.

We want to stress that another reason of interest of the GME of Kenkre, Montroll and Shlesinger is that it is formally identical to the master equation that the authors of Ref.\ \cite{mazza} derived from a quantum mechanical tight-binding model, with an erratic distribution of energy sites. The calculation was done along the lines established by Zwanzig \cite{zwanzig} using a projection operator similar to that used by Kenkre in an earlier work \cite{kenkre2}. This means that in principle understanding the connection between a master equation and CTRW might contribute to shedding light on the intriguing issue of unravelling quantum mechanical master equations. This problem becomes challenging in the case of non-Markovian processes. For some very recent references on this issue, we refer the reader  to  Refs.\ \cite{gambetta,schreiber}.  The purpose of this paper is not so ambitious as to afford direct contributions to the settlement of this problem. However, we focus our attention on the case of a two-site system that seems to be closer to the problem of decoherence of a q-bit than to the random walk over an infinite path for which the CTRW theory was originally designed. We address a problem that is closely connected to that recently discussed by Sokolov and Metzler \cite{sokolovandmetzler}, the discussion of the connection between the CTRW and a two-state non-markovian master equation. We do not discuss, however, the connection with the fractional calculus, an interesting issue addressed by the authors of Ref.\ \cite{sokolovandmetzler}. Rather we focus on the problem of how to make both the CTRW and the equivalent GME compatible with the stationary condition and the Onsager principle. For instance, we define as GME the transport equation found by the authors of Ref.\ \cite{mazza}, because it has the same structure as that proposed by Kenkre, Montroll and Shlesinger \cite{kenkre}. 
However, the quantum transport equation of the type of Ref.\ \cite{mazza} might be non stationary, due to the fact it refers to an initial condition with the system totally disentangled from the bath. In this paper we shall show how to build up a genuinely stationary GME.\\
\\
The Onsager principle is a property that makes it possible for us to derive the correlation function from the GME. If we set a condition of total equivalence between the GME and the CTRW, we find the apparently disconcerting result that only the Poisson statistics is compatible with the equivalence condition. The reason for this restriction is due to the fact that the departure from the Poisson statistics  generates memory properties
that make the GME incompatible with the Markov approximation. This means that the structure of the GME is dictated by the initial condition. If this is not stationary, the resulting GME is  not a \emph{bona fide} transport equation \cite{fox}. On the other hand, if the waiting time distribution
$\psi(t)$ is not exponential, there are aging effects. This means that we have to leave the system age till it reaches the condition where the Onsager  principle holds true. In this condition it is possible to establish a \emph{bona fide} GME. We do it, and in so doing we establish a complete equivalence between CTRW and GME. At same time, we prove what it is probably the most important result of this paper. This is that the transition from the Poisson to the non-Poisson statistics has the effect of creating a memory condition incompatible with the Markov approximation, at least in principle.  In fact, the Markov approximation, as we shall see, means that the bath is infinitely fast, and this condition, in turn, is proven to force the waiting time distribution $\psi(t)$ to be exponential. This means that the Markov condition is incompatible with  the non-Poissonian nature of the waiting time distribution.

\section{The Poisson case and the Onsager Principle}

In this section we prove that apparently the equivalence between the CTRW of Montroll and Weiss and the GME is complete only in the Poisson case. 
Let us express the GME under the following concise form: 

\begin{equation}\label{kenkre}
\frac{d}{dt} {\bf   p}(t) =  - \int_{0}^{t}  \Phi(t-t') {\it   K} {\bf   p}(t') dt',
\end{equation}
where ${\bf p}$ is  the $m$-dimensional population vector of $m$ sites, ${\it K}$ a transition matrix between the sites and $\Phi(t)$ the memory kernel.
The prescription of the CTRW \cite{montrollweiss}  yields:

\begin{equation}\label{montrolweiss}
{\bf  p}(t) = \sum_{n = 0}^{\infty} \int_{0}^{t} dt' \psi_{n}(t') \Psi(t-t') {\it  M}^{n} {\bf  p}(0).
\end{equation}
Note that $\psi_{n}(t)$ is the probability that $n$ jumps occurred and that the last took place at time $t = t'$.  This means that
\begin{equation}
\label{njumps}
\psi_{n}(t) = \int_{0}^{t} \psi_{n-1}(t-t') \psi_{1}(t')dt'  ,
\end{equation}
with
\begin{equation}
\label{zero}
\psi_{0}(t) = \delta(t).
\end{equation}
While  $ {\it M}$ is the transition matrix connecting the sites
after one jump has occurred.
The function $\Psi(t)$ is the probability that no jump occurs in the time interval $t$, namely,
\begin{equation}
\label{nojump}
\Psi(t) = \int_{t}^{\infty} \psi_{1}(t')dt'.
\end{equation}
It is evident that the time convolution structure of both Eq.\ (\ref{kenkre}) and Eq. (\ref{montrolweiss}) makes it straigthforward to derive in both cases the Laplace transform of ${\bf p}(t)$. Furthermore, Eq. (\ref{njumps}) and Eq. (\ref{nojump}) makes it possible to express the Laplace transform of ${\bf p}(t)$, in the second case, in terms of the function $\psi_{1}(t)$ only. This function is the earlier mentioned waiting time distribution, and from now on it will be denoted with the symbol $\psi(t)$. 
By comparing the Laplace transform of the GME  to the Laplace transform of the CTRW  we get

\begin{equation}\label{masterequationoutofequilibrium}
\hat \Phi(u) = \frac{u \hat \psi(u)}{1 - \hat \psi(u)} \frac{ {\it  M -I }}{{\it  K}},
\end{equation}
where ${\it I}$ is the unity operator.
We limit our discussion in this paper to the two-state case, where
\begin{equation}\label{firsttime}
{\it M}=\left(
\begin{array}{cc}
0 & 1\\
1 & 0\\
\end{array}\right)
\end{equation}
and
\begin{equation}\label{secondtime}
{\it K}=
\left(
\begin{array}{cc}
1 & -1\\
-1 & 1\\
\end{array}\right),
\end{equation}
thereby turning Eq.\ (\ref{masterequationoutofequilibrium}) into 
\begin{equation}\label{simplified}
\hat \Phi(u) = \frac{u  \hat \psi(u)}{1 - \hat \psi(u)}.
\end{equation}

We note that the two-state master equation studied in this paper has to be related to the two-site CTRW. If we assign the value $W$ to the right site, and the value $-W$ to the left site, and we adopt a discrete time representation, the motion of the random walker corresponds to a symbolic sequence $\{\xi\}$, with the form $\{W W W W W W W W - W -W -W -W W W W W W W -W -W -W -W -W -W-W...\}$, which shows a significant time persistence of both states. The waiting time distribution $\psi(t)$ is the distribution of the patches filled with either $W$'s or  $-W$'s. We assume a symmetric condition. In the case when the first moment of $\psi(t)$ is finite, the process is stationary \cite{renewal,renewaltheory} thereby making it possible to define the correlation function of the fluctuation $\xi(t)$, denoted with the symbol $\Phi_{\xi}(t)$.

 It is now the right time to use the Onsager principle. We set the Onsager principle in the form
\begin{equation}\label{realizationofonsager}
\Phi_{\xi}(t) = \frac{p_{1} (t) - p_{2}(t) } {p_{1} (0) - p_{2}(0)},
\end{equation}
where $p_{1}(t)$ and $p_{2}(t)$ are the probabilities for the random walker to be, at time $t$, in the first and second state, respectively. 
In his original work Onsager referred himself to the case of a macroscopic fluctuation that is supposed to regress to the vanishing equilibrium value through a phenomemological equation of motion. We realize this macroscopic fluctuation selecting a large number $N$ of walkers, divided into two groups, with $N_{1}$ and $N_{2}$ walkers  belonging to the first and second state, respectively. Then we relate this choice to the probabilities 
$p_{1}(t)$ and $p_{2}(t)$, by setting $p_{1} = N_{1}/N$ and $p_{2} = N_{2}/N$. The regression to equilibrium of this macroscopic fluctuation does not fit any phenomenological  law. For this reason, we can refer ourselves to Eq.\ (\ref{realizationofonsager}) as a generalized version of the Onsager principle. We plan to derive the mathematical expression of the law of regression to equilibrium 
from the use of this generalized version of the Onsager's regression principle, which, as shown in Section 3, is exact in the dichotomous case.  
Due to its nature, the GME can afford information only on $ p_{1}(t)$ and $ p_{2}(t)$. There is no direct information on the correlation function $\Phi_{\xi}(t)$, and the only possible way is through  the Onsager's principle, which, as shown by Eq.\ (\ref{realizationofonsager}), implies that an initial off-equilibrium condition, $p_{1} (0) - p_{2}(0) \neq 0$, is established. Then, the regression to equilibrium is expected to take place in such a way as to establish a connection with the correlation function in the form of Eq.\ (\ref{realizationofonsager}).  
This assumption, applied to Eq.\ (\ref{kenkre}), yields:
\begin{equation}\label{timeconvoluted}
\frac{d}{dt} \Phi_{\xi}(t) = - 2\int_{0}^{t}dt' \Phi_{\xi}(t- t')\Phi(t').
\end{equation}
This equation is the non-Markovian counterpart of the phenomenological regression to equilibrium of the original work of Onsager \cite{onsager}. Using the time convoluted expression of Eq.\ (\ref{timeconvoluted}) and the central result of Eq.\ (\ref{simplified}), we establish a connection between the correlation function $\Phi_{\xi}(t)$ and the waiting time distribution $\psi(t)$, through their Laplace transforms, as follows:
\begin{equation}\label{firstconnection}
\hat \Phi_{\xi}(u)  = \frac{1}{u + \frac{2 u \hat \psi(u)}{(1 - \hat \psi(u))}} .
\end{equation}

On the other hand, the correlation function is connected to $\psi(t)$ through properties established by the renewal theory \cite{renewal}. 
To properly establish this connection, we have to notice, with Zumofen and Klafter \cite{klafter}, that it is convenient to introduce another  type of waiting time distribution that we call $\psi^{*}(t)$. What's the connection between $\psi(t)$ and  $\psi^{*}(t)$? The waiting time distribution $\psi(t)$ is the experimental waiting time distributon. It could be evaluated experimentally observing the sequence $\{\xi\}$, and recording the time length of the laminar regions occupied only by $W$ or by $-W$. However, we can imagine that a theoretical waiting time distribution exists, denoted by $\psi^{*}(t)$, and that the sequence $\xi(t)$ is obtained as follows. We select randomly a number $t_{1}$ from the distribution $\psi^{*}(t)$. We toss a coin, and assign to the first laminar region, of length $t_{1}$, the symbol $W$ or $-W$, according to the coin tossing prescription. At the end of this laminar region, first we select, again from the distribution $\psi^{*}(t)$, a number $t_{2}$. This is the length of the second laminar region. We toss the coin again to decide the sign of it. It is evident that there is $50\% $ probability of getting the same sign as the earlier laminar region. We proceed in the same way with the length and the sign of third
laminar region, and so on. We adopt this rule to create the sequence $\{\xi(t)\}$. Thus, from the
renewal theory \cite{renewal} we obtain the following important result:
\begin{equation}\label{geiselcrucial1}
 \Phi_{\xi}(t) = \frac{1}{\langle t \rangle} \int_{t}^{\infty} (t'-t) \psi^{*}(t')dt',
\end{equation}
where $\langle t \rangle$ is the mean waiting time of the $\psi^{*}(t)$-distribution.  It is interesting to notice that this equation means that the second derivative of the correlation function is proportional to $\psi^{*}(t)$,
\begin{equation}\label{geiselcrucial2}
\frac{d^{2}}{dt^{2}} \Phi_{\xi}(t) = \frac{\psi^{*}(t)}{\langle t \rangle}.
\end{equation}

Zumofen and Klafter \cite{klafter} explained with clear physical arguments the connection between $\psi(t)$ and $\psi^{*}(t)$.  
They prove that their Laplace transforms are related one to the other by
\begin{equation}\label{connection}
\hat \psi^{*}(u) = \frac{2 \hat \psi(u)} {1 + \hat \psi(u)}.
\end{equation}
Using Eq.\ (\ref{geiselcrucial2}) and Eq.\ (\ref{connection}) we derive a further connection between the Laplace transform of $\Phi_{\xi}(t)$ and the Laplace transform of $\psi(t)$. In fact, by Laplace transforming Eq.\ (\ref{geiselcrucial2}) and using Eq.\ (\ref{connection}), we derive a new expression for the Laplace transform of $\Phi_{\xi}(t)$. By equating this new expression to that of Eq.\ (\ref{firstconnection}), we find, after some algebra, the following form for the Laplace transform of $\psi(t)$,
\begin{equation}\label{awful}
\hat \psi(u) = \frac{- \frac{1}{2} \Phi_{\xi}^{'}(0) }{u + \frac{1}{\langle t \rangle }  + \frac{1}{2} \Phi_{\xi}^{'}(0)}.
\end{equation}
Using Eq.\ (\ref{geiselcrucial1}) we prove that
\begin{equation}
\Phi_{\xi}^{'}(0) = - \frac{1}{\langle t \rangle}.
\end{equation}
This equation allows us to  write Eq.\ (\ref{awful}) under the form
\begin{equation}
\hat \psi(u) = \frac{1}{2 \langle t \rangle } \frac {1 }{(u + \frac{1}{2 \langle t \rangle }) }. 
\end{equation}
This means that $\psi(t)$ is exponential and the explicit forms of $\psi(t)$ and $\Phi_{\xi}(t)$ are given by
\begin{equation}
\label{possiblecondition1}
\psi(t) = \frac{\gamma}{2} \exp(-\frac{\gamma}{2} t)
\end{equation}
and
\begin{equation}
\label{possiblecondition2}
\Phi_{\xi}(t) = \exp(-\gamma t),
\end{equation}
respectively. 

This is the first important result of this paper. It is to some extent disconcerting since it seems to restrict the complete equivalence between GME and CTRW to the exponential case, with the effect of making the GME useless, since in the exponential case, it turns out to be an ordinary, memory-less, master equation. We want to notice that Sokolov and Metzler \cite{sokolovandmetzler} have recently discussed the derivation of a fractional transport equation totally equivalent to the CTRW. However, they did not discuss the intimately related problem of the formal equivalence between CTRW and GME, which is the main goal of the present paper. 

The second important result of this paper is the discovery of a GME that is equivalent to the CTRW, with no restriction to the Poisson case.
 This second important result will be illustrasted in Section 5. To derive this important result we have to mention a fact recently observed by Barkai \cite{barkai}: the Montroll-Weiss CTRW refers to a non-stationary condition.
This means that the Onsager's regression principle is invalidated. Therefore, we shall conclude that the disconcerting result of this section is  due to the fact that the case of Poisson statistics is the only one where aging does not exist.   At this stage, this is only a conjecture. We can notice, however, that this is a plausible conjecture. In fact, the exponential condition of Eq.\ (\ref{possiblecondition1}) when applied to Eq.\ (\ref{simplified}) makes $\hat \Phi(u)$
independent of $u$, thereby implying that instantaneous regression to the vanishing value of the memory kernel $\Phi(t)$. This means that the fluctuations of the variable $\xi$ take place with its bath at equilibrium.

We give a further support to the conjecture that aging is the reason of the apparently disconcerting result of this section by referring ourselves to the work of Ref.\ \cite{mazza}. The authors of this paper found a transport equation with the same form as that of Eq.\ (\ref{kenkre}) , with no inhomogeneous term.
The adoption of the Zwanzig projection method makes this interesting result possible, provided that special initial conditions are chosen. These conditions might depart from the entanglement between the system of interest and its bath, necessary to ensure the stationary condition. On the other hand, the Hamiltonian treatment implies that the system exerts an action on its bath. Thus, the choice of an initial condition annihilating the inhomogeneous term might have the effect of perturbing the bath, with an ensuing  long lasting re-equilibration  process, and so a persistent non-stationary condition.

\section{Onsager's regression for dichotomous signals}

We note that the constraint posed in the earlier section on the waiting time distribution $\psi(t)$,  to ensure the equivalence  between the Master equation and the CTRW,  has been based on the assumption that the Onsager principle is valid. 
The fact that the equivalence between CTRW and GME is restricted only to the Poissonian case, and so to the case of ordinary master equation, might generate the false impression that the Onsager principle is valid only in the Poissonian case. It is not so. It is already known \cite{zambon} that in the case of Gaussian statistics the Onsager's regression hypothesis is  exact, and it holds true for initial excitation of any intensity. 
In this section, having in mind the two-state master equation, we focus our attention on the case of dichotomous statistics. In the dichotomous case, we show that  the Onsager's regression hypothesis turns out to be exact again,  for initial departures from equilibrium of whatsoever intensity, and with any type of relaxation process, with an exponential form, and with an inverse power law form as well, provided that the system is ergodic. In short, in this section we prove that in the dichotomous case the Onsager regression principle holds true provided that the system is ergodic.

The ergodic property implies that we can create a Gibbs ensemble in two different, but equivalent, ways. The first refers to the original idea of Gibbs, that the same system is repeated infinitely many times. This means that we have to generate infinitely many sequences using the same physical prescription. 
The second way is based on the adoption of only one sequence, denoted by $\xi(t)$, which is supposed to be infinitely long. Then we define a generic trajectory $\xi^{(s)}(t)$ with the following prescription
\begin{equation}
\label{manytrajectoriesfromone}
\xi^{(s)}(t) = \xi(t +s).
\end{equation}
In the practical case of a numerical treatment, time is discrete, time unity being, for instance $t = 1$. In this case the superscript $s$ is an integer number. 

 We denote by $P_{+}$ and $P_{-}$ the probability that the variable $\xi$ gets the value $W$ and $-W$, respectively. These probabilities are frequencies that can be evaluated using either the first or the second ensemble of sequences. We divide our ensemble of sequences into two groups,  characterized by the the initial condition $\xi(0)=1$,
and $\xi(0)=-1$, respectively.    We define two distinct averages
for the two groups, denoted by $\langle \cdot \rangle_{+} $ and
 $\langle \cdot \rangle_-$, respectively. Of course, we have that:

\begin{equation}\label{defaverage}
\langle \cdot \rangle = 
P_{+} \langle \cdot \rangle_{+} + P_{-} \langle \cdot \rangle_{-},
\end{equation}
with
\begin{equation}
P_{+} + P_{-} =1.
\end{equation}

Although it is not essential for the main goal of this paper, we make our discussion as general as possible. Thus, we do not assume the two probabilities to be equal. We do not rule out that a bias might exists, given by
\begin{equation}
P_{+} - P_{-} = c.
\end{equation}
In the ergodic, and stationary, condition that we are assuming, the stationary autocorrelation function exists, and it is defined as
\begin{equation}\label{correlationfunction1}
\Phi_{\xi}(t) \equiv \frac{\langle \xi(0) \langle \xi \rangle \rangle 
\langle \xi(t) \langle \xi \rangle \rangle}
{\langle \xi ^2 \rangle - \langle \xi \rangle ^2}
= \frac{\langle \xi(0) \xi(t) \rangle -c^2}{1-c^2}.
\end{equation}
Expressing the total average in terms of the averages over the two groups, according to the prescription of Eq.\ (\ref{defaverage}), we make Eq.\ (\ref{correlationfunction1}) become
\begin{eqnarray}\label{correlationfunction2}
%\begin{split}
\Phi_{\xi}(t) & =& \frac{\langle (+1) \xi(t) \rangle_{+} 
+ \langle (-1) \xi(t) \rangle_{-} -c^2}{1-c^2}\nonumber \\&=&
\frac{\langle \xi(t) \rangle_{+} - \langle \xi(t) \rangle_{-} -c^2}{1-c^2}.
%\end{split}
\end{eqnarray}
Note that on the same token  the property $ \langle \xi \rangle =c $ can be written  as follows
\begin{equation}\label{firstproperty}
P_{+} \langle \xi \rangle_{+} + P_{-} \langle \xi \rangle_{-} =c. 
\end{equation}
We also note that 
\begin{equation}\label{secondproperty}
P_{+}=\frac{c+1}{2} .
\end{equation}
Using Eqs.(\ref{firstproperty}) and (\ref{secondproperty})
we express the correlation function of Eq.\ (\ref{correlationfunction2}) as follows:
\begin{equation}\label{better}
\Phi_{\xi}(t)=\frac{2 P_{+} \langle \xi \rangle_{+} -c -c^2}{1-c^2}
= \frac{\langle \xi \rangle_{+} (c+1) -c (c+1)}{1-c^2}.
\end{equation}
Finally, by dividing both numerator and denominator of the ratio corresponding to the last term of Eq.\ (\ref{better}) by $ 1+c $, we obtain
\begin{equation}\label{exactonsager}
\Phi_{\xi}(t)=\frac{\langle \xi \rangle_{+} -c}{1-c}.
\end{equation}
Eq.\ (\ref{exactonsager}) can be interpreted as follows. The group of trajectories corresponding to  the condition $\xi(0)=1$ can be thought of as a way of creating an out of equilibrium condition, $ \langle \xi \rangle_{+}$. This out of equilibrium condition undergoes a process of regression to equilibrium that is proportional to the equilibrium correlation function. Using Eq.\ (\ref{firstproperty}), we can express the correlation function in terms of $\langle \xi\rangle_{-}$, and we reach the same conclusion. It is evident that the same conclusion would be reached using an arbitrary mixture of $\langle \xi \rangle_{+}$ and $\langle \xi \rangle_{-}$, departing from the vanishing equilibrium value. In conclusion, the Onsager's regression principle is fulfilled.

We notice that the waiting time distributions that we are considering in this paper have an inverse power law nature with the power index $\mu$. If this power index fits the condition $\mu >  2$ the ergodic condition is ensured. We shall focus our attention on this condition, and we shall prove that we can make the CTRW  compatible with the GME if we set the constraint that the system is aged enough as to make the Onsager principle valid. As we shall see, this means the adoption of a form of CTRW different from that of Montroll and Weiss \cite{montrollweiss},  which corresponds  to a condition very far from the stationary state. It is evident that the ergodic condition might become hard to fulfill in practice, with relaxation processes described by inverse power laws. This is in fact the case when aging becomes important.

\section{Aging in renewal processes}

In Section 2 we found that the GME and the CTRW are equivalent only in the Poisson case. Since the use of the Onsager's regression assumption is the key ingredient used to establish this equivalence, we might be tempted to conclude that the Onsager principle does not hold true in the non-Possonian case. In Section 3, we found that it is not so, and that the Onsager principle is valid, provided that $\mu >2$. 
Here we shall prove that the non-Poisson case results in aging effects, which has significant consequences also in the case $\mu > 2$. If the CTRW used does not refer to the stationary (aged) condition, it cannot be compatible with the Onsager regression principle.  This is the second result of this paper. The problem of aging within the context of intermittent processes has been discussed recently in a very attractive paper by Barkai \cite{barkai}. The work of Barkai focuses on the 
non-stationary condition, $\mu < 2$, where no invariant measure exists, and aging lasts forever. In this case the Onsager regression principle is violated for reasons explicitly pointed out by Onsager himself. In fact, Onsager repeatedly mentioned \cite{onsager} that his principle has to be applied to aged systems. The condition of aged systems implies a state of thermodynamic equilibrium. In the case $\mu < 2$ equilibrium is never reached, not even asymptotically. Thus, in this case the Onsager's principle is always violated. In Section 4A, using   the renewal theory \cite{renewaltheory}, we shall study the case $\mu >2$, and we shall discuss two limiting conditions, corresponding to the birth and the death of the system, respectively. It is interesting to remark that death, meaning an ordinary thermodynamical condition, takes place eventually at the end of an infinitely long aging process.

\subsection{ Young and aged systems}

First of all,  we should define,  as Barkai does \cite{barkai},  the waiting time distribution
$\psi_{t_{a}} (t)$.  This means that we establish non-stationary conditions at time $t_{a} < 0$, and we begin our observation at time $t = 0$. The waiting time distribution depends on $t_{a}$. 
The naive conviction that the waiting time distribution is given by $\psi(t)$, actually is based on the assumption that $t_{a} = 0$. This means, in fact, that we are considering a set of random walkers and that at time $t = 0$ all of them begin their sojourn in the laminar region. This leads immediately to $\psi_{t_{a} = 0}(t)= \psi(t)$. It is straighforward to evaluate also the distribution  of sojourn times corresponding to $t_{a} = - \infty$. We denote this distribution with the symbol $\psi_{\infty}(t)$. 

This is the stationary case, corresponding to the following procedure.  In the stationary condition, the probability of selecting 
 a laminar zone of length $T$, by a random choice,  is $\frac{T}{\langle \tau \rangle} \psi(T) dT$, where $\langle \tau \rangle $ is the mean length of a laminar zone.
The  probability density of observing the first change of laminar phase  after a time $t$, being  in a laminar
zone of length $T$, is $\Theta(T-t) \frac{1}{T}$. Consequently for the probability density of having the first change of laminar phase at time $t$ (i.e. $ \psi_{\infty}(t)$ ), we have, integrating over all possible $T$:
\begin{eqnarray}
\label{infinitelyoldpsi}
%\begin{split}
\psi_{\infty}(t)&=& \frac{1}{\langle \tau \rangle} 
\int\limits_{0}^{\infty} {\rm d}T T \psi(T) \frac{1}{T} \Theta(T-t)\nonumber\\ 
 &=&\frac{1}{\langle \tau \rangle } \int\limits_{t}^{\infty} {\rm d}T \psi(T)  .
%\end{split}
\end{eqnarray}

Note that due to the renewal theory, the distribution $\psi_{\infty}(t)$ concerns only  the time that we have to wait to detect the first event.   
After the first event we have a total rejuvination. In fact, measuring the time at the moment of the first jump is equivalent to beginning  the measurement process at the precise moment when the walker enters the laminar region. Thus the Laplace transform of ${\bf p}(t)$ is given by
\begin{equation}\label{oldage}
{\bf \hat p}(u)=( \hat\Psi_{\infty}(u){\it I} +\frac{\hat\psi_{\infty}(u)\hat\Psi(u) {\it M}}{1-\hat\psi(u) {\it M}}){\bf p}(0).
\end{equation}
This result is derived from  Eq.\ (\ref{montrolweiss}) as follows. First of all, we evaluate the Laplace transform of Eq.\ (\ref{montrolweiss}). We obtain
\begin{eqnarray*}
\label{young}
%\begin{split}
{\bf \hat p}(u)&=&  (\hat \Psi(u) + \hat \Psi(u)  \hat  \psi(u)  {\it M} + \hat \Psi(u)  \hat \psi^2 {\it M}^2 + ... ) {\bf p}(0) \nonumber\\
&=& \hat \Psi(u)  \sum\limits_{n=0}^{\infty} ( \hat \psi {\it M} )^n {\bf p}(0)
= \frac{\hat \Psi(u) }{1-\hat \psi(u) {\it M} } {\bf p}(0).
%\end{split}
\end{eqnarray*}
Then, we replace the probability of occurrence, or of non occurrence, of the first event, called $\psi(t)$ and  $\Psi(t)$, respectively, with the corresponding aged quantities. These are called
$\psi_{\infty}(t)$ and $\Psi_{\infty}(t)$, respectively. The function $\psi_{\infty}(t)$ is given by Eq.\ (\ref{infinitelyoldpsi}) and the function $\Psi_{\infty}(t)$ is given by
\begin{equation}\label{agedPsi}
\Psi_{\infty}(t) = \int_{t}^{\infty} \psi_{\infty}(t') dt'.
\end{equation}
All this yields Eq.\ (\ref{oldage}). 

	We expect that the inverse Laplace transform of Eq.\ (\ref{young}),
 is a  function of $t$ that asymptotically
will become equivalent to the asymptotic value of the inverse Laplace transform of Eq.(\ref{oldage}). 
What about the Onsager
prescription of Eq.\ (\ref{realizationofonsager})?  If we adopt the GME corresponding to the young condition, the adoption of Eq.\ (\ref{realizationofonsager}) would be equivalent, as far as the left-hand-side of this equation is concerned, to adopting the non-stationary condition. We have to remark, in fact, that the GME corresponding to the Montroll-Weiss CTRW is not a \emph{bona fide} master equation. The importance of a \emph{bona fide} transport equation has been pointed out in a remarkable paper by Fox \cite{fox}. 
A master equation 
 is a \emph{bona fide} master equation when it can be used with any initial condition  (like the \emph{bona fide} Fokker-Planck equation of Ref.\ \cite{fox}) .  The master equation of Section 3, on the contrary, implies the choice of only one initial condition. The next subsection is devoted to sheding more light into this important aspect.

\subsection{A theory for systems of any age}
The purpose of this subsection is to shed further light into the aging problem, and into the ensuing conflict with the Onsager postulate. At the same time we shall derive an expression for the distribution of the first exit times valid for any age, and not only for $t_{a} = 0$ and $t_{a} = - \infty$, which denote observation taking place at the moment of birth and death of the dynamic process under study, respectively. 
This discussion is based on the following dynamic model. A variable $x$ moves in the interval $I \equiv [0,1]$ according to the following prescription
\begin{equation}
\frac{dx}{dt} = \lambda x(t)^{z}.
\end{equation} 
When it reaches the point $x = 1$ it is injected back with uniform distribution, thereby producing another laminar region. We assign   alternated signs to the sequel of laminar regions. The sojourn in a laminar region with a given sign is equivalent to sojourn in one of the two states  discussed in Section 2. The resulting waiting time distribution coincides with the one earlier called $\psi(t)$. The explicit expression for $\psi(t)$ is obtained by expressing the exit time $t$ as a function of the initial condition $x_{0}$. Then we have to assume
\begin{equation}
p(x_{0}) dx_{0} = \psi(t)dt .
\end{equation}
The choice of a uniform of back injection process implies $p(x_{0}) = 1$. 
This is the condition behind the CTRW of Montroll and Weiss.  All this results into
\begin{equation}
\psi(t) = (\mu -1) \frac{T^{(\mu -1)}}{(T + t)^{\mu}},
\end{equation}
with
\begin{equation}
\mu = \frac{z}{(z-1)}
\end{equation}
and
\begin{equation}
T = \frac{1}{\lambda (z-1)}.
\end{equation}

We propose for the aging process a calculation procedure different from that adopted by Barkai \cite{barkai}. As we shall see, this procedure  yields the same results as those of Barkai. As done in Ref.\ \cite{rosa},  we discuss the 
equation of motion for the probability density $\rho(x,t)$. 
This is given by
\begin{equation}
\label{cheerfulequation}
\frac{\partial}{\partial t} \rho(x,t) = - \frac{\partial}{\partial x} ( \lambda x^{z} \rho(x,t)) + C(t) .
\end{equation}

The time evolution of the distribution function is
described by the following formula
\begin{equation}\label{evol}
\rho(x, t) \simeq
\int_{-t_{a}}^{t}\frac{C(\tau+t_{a})}{[1+1/\alpha(t-\tau)x^{1/\alpha}]^{1+\alpha}}d\tau,
\end{equation}
where we have used the following new parameter:
\begin{equation}
\label{newparameter}
\alpha = \frac{1}{(z-1)}.
\end{equation}

We have assumed that the flat initial distribution is assigned at
time $t=-t_{a}$. We have neglected the contribution $[1 + 1/\alpha(t+t_{a})^{(1/\alpha)}]^{-(1+\alpha)}$, necessary to recover at $t = -t_{a}$, the flat initial distribution. On the basis of the approach that we shall detail hereby to derive the distribution of the first exit times,  it is straightforward to check  that this term yields negligible contributions. 
Using the property
\begin{equation}
\label{wellknownprescription}
\rho(x, t = 0) dx = \psi_{t_{a}}(t) dt,
\end{equation}
where $\rho(x, t=0)$ is given by Eq.\ (\ref{evol}), with $t=0$,
we can evaluate the distribution function of the first exit times in general, and not only for $t_{a} = 0$ and $t_{a} = - \infty$.
It is interesting to remark that the infinitely old distribution $\psi_{\infty}(t)$ of Section 4 A, is easily obtained by noticing that
\begin{equation}
\label{infinitelyolddistribution}
\rho(x, \infty) = (2-z)/ x^{z-1}.
\end{equation}
Using Eqs.(\ref{infinitelyolddistribution}) and (\ref{wellknownprescription}), after some algebra, we rederive the distribution of Eq.\ (\ref{infinitelyoldpsi}). 

We can also establish a connection with the case discussed by Barkai \cite{barkai}, namely the case $z > 2$. In the case of $z>2 \Rightarrow \alpha<1$,
the long time behavior of the function $C(t)$ is
\begin{equation}\label{longtime}
C(t) \simeq
\frac{\sin(\alpha\pi)}{\alpha^{\alpha}\pi}t^{-(1-\alpha)}.
\end{equation}
We now use the  prescription of Eq.\ (\ref{wellknownprescription}).
A close inspection of Eqs. (\ref{evol}) and (\ref{longtime})
reveals that it is not possible to give a close formula for the
function $\psi_{t_{a}}(t)$. Then, we must study the two cases $t_{a} \gg
t$ and $t_{a} \ll t$, separately.
\begin{itemize}
\item $t_{a} \gg t$:
\begin{eqnarray}\label{little_ta}
\rho(x, t=0) & \sim &
\int_{-t_{a}}^{0}\frac{(\tau+t_{a})^{-(1-\alpha)}}{[1-1/\alpha\tau
x^{1/\alpha}]^{1+\alpha}}d\tau \nonumber\\ & = &
t_{a}^{-1}\int_{-1}^{0}\frac{(1+y)^{-(1-\alpha)}}{[1/t_{a}-y/\alpha
x^{1/\alpha}]^{1+\alpha}}dy \nonumber\\ & = &
t_{a}^{-1}\int_{0}^{1}\frac{(1-y)^{-(1-\alpha)}}{[1/t_{a}+y/\alpha
x^{1/\alpha}]^{1+\alpha}}dy \nonumber\\ & \sim &
t_{a}^{-1}\int_{0}^{1}\frac{1}{[1/t_{a}+y/\alpha
x^{1/\alpha}]^{1+\alpha}}dy \nonumber\\ & = &
t_{a}^{\alpha}\int_{0}^{1}\frac{1}{[1+y/\alpha t_{a}
x^{1/\alpha}]^{1+\alpha}}dy \nonumber\\ & \sim &
\frac{t_{a}^{-(1-\alpha)}}{x^{1/\alpha}}.
\end{eqnarray}
\end{itemize}
In Eq.\ (\ref{little_ta}), we have carried out the operation   $\lim_{t_{a}
\rightarrow \infty}$, while keeping  $x \neq 0$ fixed. The next step is based on the use of  Eq.
(\ref{wellknownprescription}), of  $x =
(1+t/\alpha)^{-\alpha}$ and of  $\left|\frac{dx}{dt}\right| =
(1+t/\alpha)^{-(\alpha+1)}$. This allows us to obtain
\begin{equation}\label{first_h}
\left.\psi_{t_{a}}(t)\right|_{t_{a} \gg t} \sim
\frac{t_{a}^{-(1-\alpha)}}{t^{\alpha}}.
\end{equation}
\begin{itemize}
\item $t_{a} \ll t$:
\begin{eqnarray}\label{little_ta1}
\rho(x, t=0) & \sim &
\int_{-t_{a}}^{0}\frac{(\tau+t_{a})^{-(1-\alpha)}}{[1-1/\alpha\tau
x^{1/\alpha}]^{1+\alpha}}d\tau \nonumber\\ & \sim &
\int_{-t_{a}}^{0}(\tau+t_{a})^{-(1-\alpha)}d\tau \nonumber\\ & = &
t_{a}^{\alpha}.
\end{eqnarray}
Thus, in this condition we have
\begin{equation}\label{second_h}
\left.\psi_{t_{a}}(t)\right|_{t_{a} \ll t} \sim
\frac{t_{a}^{\alpha}}{t^{1+\alpha}}.
\end{equation}
\end{itemize}
Let us note that  Eqs. (\ref{first_h}) and (\ref{second_h})
correctly reproduce the behavior predicted by  Ref.\ \cite{barkai})
in the corresponding limits by means of the formula:
\begin{equation}\label{barkai}
\psi_{t_{a}}(t) \sim
\frac{\sin(\alpha\pi)}{\pi}\frac{t_{a}^{\alpha}}{t^{\alpha}(t+t_{a})},
\end{equation}
where $\alpha \equiv \frac{1}{z-1}$.
In conclusion, our way of proceeding fits the conclusions of Barkai \cite{barkai}.  As earlier pointed out, in the condition $\mu < 2$ the Onsager principle is always violated. This is expected, since it is known \cite{bettin} that the non-stationary condition $\mu < 2$ is incompatible with the existence of the stationary correlation function $\Phi_{\xi}(t)$.

\section {The Stationary Master Equation}
In this section we prove that the GME can be made equivalent to the 
CTRW, with no restriction to the Poisson statistics, if we set 
 the stationary condition in both of them.
The deep difference between stationary and non-stationary condition has been discussed years ago by Zumofen and Klafter in two pioneering  papers \cite{a,d}. This means that we have to compare the GME of Eq.\ (\ref{kenkre}) to the stationary CTRW, which yields

\begin{equation}
\label{CTRW}
{\bf \hat p}(u)=( \hat\Psi_{\infty}(u){\it I} +\frac{\hat\psi_{\infty}(u)\hat\Psi(u) {\it M}}{1-\hat\psi(u) {\it M}}){\bf p}(0),
\end{equation}
where, according to the stationary prescription \cite{chemphys}:
\begin{equation}
\label{agedpsi}
\psi_{\infty}(t)  = \frac{1}{\langle \tau \rangle} \int_{t}^{\infty} dt' \psi(t'),
\end{equation}
 as also explained in Section 4A, and  $\langle \tau \rangle$ is used again  to denote the mean waiting time of the distribution $\psi(t)$.
For ${\it M}$ we make the same choice as in Eq.\ (\ref{firsttime}). For calculation convenience, we adopt  for ${\it K}$ the more general symmetric form:
\begin{equation}
{\it K}=\left(
\begin{array}{cc}
x & y\\
y & z\\
\end{array}\right).
\end{equation}
From (\ref{kenkre}) we get:

\begin{equation}\label{LtMeq}
{\bf \hat p}(u)=\frac{1}{u {\it I}+ \hat\Phi(u) {\it K}} {\bf p}(0).
\end{equation}

By equating (\ref{LtMeq}) and (\ref{CTRW}) we derive

\begin{equation}\label{MeqCTRW}
\frac{1}{u {\it I}+ \hat\Phi(u) {\it K}}=\hat\Psi_{\infty}(u){\it I} +\hat\psi_{\infty}(u)\hat\Psi(u) {\it M}\sum_{n=0}^{\infty}\hat\psi(u)^n {\it M}^{n}.
\end{equation}
By multiplying both sides by $ {u {\it I}+ \hat\Phi(u) {\it K}}$ and using the properties ${\it M}^{2n}={\it I} $ and ${\it M}^{2n+1}={\it M}$, we obtain:
%\begin{eqnarray}\label{calc}
\begin{equation}\label{calc}
\begin{split}
{\it I}=&\hat\Psi_{\infty}(u)(u {\it I}+\hat\Phi(u) {\it K})+ \hat\psi_{\infty}(u)\hat\Psi(u)\left( \frac{ u}{1-\hat\psi(u)^2}{\it M}  \right. \\
&\left. + \frac{\hat\Phi(u) }{1-\hat\psi(u)^2}{\it M} {\it K}+   \frac{\hat\psi(u) u }{1-\hat\psi(u)^2}{\it I} +    \frac{\hat\psi(u) \hat\Phi(u)}{1-\hat\psi(u)^2}{\it K} \right) .   
\end{split}
\end{equation}
%\end{eqnarray}
With some algebra we extract from Eq.\ (\ref{calc})
the following set of equations for $ x, y $ and $z$:
%\begin{eqnarray*}
\begin{equation*}
\begin{split}
 1=&\hat\Psi_{\infty}(u)(u+\hat\Phi(u) x)+\\& \hat\psi_{\infty}(u)  \hat\Psi(u) \left( \frac{\hat\Phi(u)}{1-\hat\psi(u)^2}y+\frac{\hat\psi(u) u}{1-\hat\psi(u)^2} +    \frac{\hat\psi(u) \hat\Phi(u)}{1-\hat\psi(u)^2} x \right)   \\
   0=&\hat\Psi_{\infty}(u)\hat\Phi(u) y+ \\& \hat\psi_{\infty}(u)  \hat\Psi(u) \left(\frac{\hat\Phi(u)}{1-\hat\psi(u)^2}z+\frac{ u}{1-\hat\psi(u)^2} +    \frac{\hat\psi(u) \hat\Phi(u)}{1-\hat\psi(u)^2} y \right)     \\
    0= &   \hat\Psi_{\infty}(u)\hat\Phi(u) y+ \\& \hat\psi_{\infty}(u)  \hat\Psi(u) \left(\frac{\hat\Phi(u)}{1-\hat\psi(u)^2}x+\frac{ u}{1-\hat\psi(u)^2} +    \frac{\hat\psi(u) \hat\Phi(u)}{1-\hat\psi(u)^2} y \right)     \\
     1= &   \hat\Psi_{\infty}(u)(u+\hat\Phi(u) z)+\\& \hat\psi_{\infty}(u)  \hat\Psi(u) \left( \frac{\hat\Phi(u)}{1-\hat\psi(u)^2}y+\frac{\hat\psi(u) u}{1-\hat\psi(u)^2} +    \frac{\hat\psi(u) \hat\Phi(u)}{1-\hat\psi(u)^2} z \right)      \\\
\end{split}
%\end{eqnarray*}
\end{equation*}
The solution of this set of equations is given by $ x = z, y = -x $ and
\begin{equation}
 x=\frac{u (1-\hat\psi(u))}{\hat\Phi(u)(-2+u \langle \tau \rangle+(2+u \langle \tau \rangle)\hat\psi(u))}   .
\end{equation}
So, the GME  and the stationary CTRW are compatible if ${\it K}$ has the same form as that of Eq.\ (\ref{secondtime})
and
\begin{equation}\label{result}
\hat\Phi(u)=\frac{u (1-\hat\psi(u))}{(-2+u \langle \tau \rangle+(2+u \langle \tau \rangle)\hat\psi(u))} .
\end{equation}
Using the Onsager's principle in the form of Eq.\ (\ref{realizationofonsager}) and consequently the time convoluted equation of Eq.\ (\ref{timeconvoluted}), we obtain

\begin{equation}
\hat\Phi_{\xi}(u)=\frac{1}{u+2 \hat\Phi(u)}.
\end{equation}

Finally, using for $\hat\Phi(u)$ the expression of Eq.\ (\ref{result}) and for $\hat\Phi_{\xi}(u)$ the expression afforded  by the renewal theory,
we get an identity in $\hat\psi(u) $ (the Laplace transform of the waiting times distribution). This means that the aged CTRW is totally equivalent to the GME with the memory kernel given by Eq.\ (\ref{result}). 

\section{Concluding remarks}
This paper affords a significant contribution to understanding the role of aging in systems departing from Poisson statistics. The Onsager regression principle rests on the second principle of thermodynamics, and it applies to aged systems. However, non-Poissonian statistics might imply a very slow regression to equilibrium, so as to create conditions, correctly depicted by the Montroll-Weiss CTRW, but incompatible with the Onsager regression postulate. This might be a condition different from the out of equilibrium thermodynamics, and represent instead a state of matter intermediate between the dynamic and the thermodynamic condition. This new condition has been recently proposed to explain the emergency of life on earth and has been denoted with the appealing term of \emph{living state of matter} \cite{livingstateofmatter}.

The master equation of Section 5 corresponds to the aged systems where the Onsager principle holds true. The Montroll-Weiss CTRW and the corresponding master equation refers to the dynamics of a young system. From a technical point of view one might wonder if it is possible to produce a master equation, or a CTRW, reflecting correctly the process of aging.
The young and the old sequence reflect two different initial conditions of the system. Moving from  the young condition the system will eventually relax to the equilibrium, but this relaxation
will not follow the correlation function behavior. 

Eventually,  the system will settle at equilibrium. However, to establish, at least ideally, the validity of the Onsager regression principle, we have to set some suitable out of equilibrium initial conditions. If we do select initial conditions that do not fit the stationary prescription \cite{a,d}, the correlation function $\Phi_{\xi}(t)$ cannot be defined, and consequently we are forced to determine the memory kernel of the GME of Eq.\ (\ref{kenkre}) using the non-stationary prescription of Eq.\ (\ref{simplified}).  Neverthless, this equation will produce an exact time evolution. Let ${\bf p}(0)$ evolve in time till it reaches the new condition ${\bf p}(T)$. If we adopt this as a new initial condition, namely we set the new origin of time at $t = T$, and we apply again Eq.\ (\ref{kenkre}) to predict the ensuing time evolution, we get a wrong prediction. In fact, the GME equivalent to the CTRW of Montroll and Weiss is not a \emph{bona fide} master equation \cite{fox}. We should build up a new GME equation with the distribution of the first jumps given by  $\psi_{t_{a}=-T} (t)$.   If, on the contrary, we select an initial out of equilibrium condition, as done in Section 3, which is compatible with the bath being at equilibrium, then the Onsager principle is fulfilled, the GME with the kernel of Eq.\ (\ref{result}) is a \emph{bona fide} master equation and can be used with any kind of initial conditions concerning ${\bf p}$.\\ 

To derive the results here illustrated, we significantly benefitted from the important work of Refs. \cite{sokolovandmetzler,barkai}. However, we want to point out that the material derivative of Ref.\ \cite{sokolovandmetzler}, although more advantageous than the GME to study the influence of an external force field, is equivalent to the GME of Ref.\ \cite{kenkre}, with the
condition of Eq.\ (\ref{masterequationoutofequilibrium}) rather than the stationary condition of Eq.\ (\ref{result}), and, consequently, as proved in this paper,it violates the Onsager principle. As to the  work of Barkai on aging \cite{barkai}, we would like to note that it is confined to the case $\mu < 2$, a condition incompatible with  the stationary master equation found in this paper, namely the master equation with the memory kernel of Eq.\ (\ref{result}). Barkai reach the conclusion that the aging effect for $\mu > 2$ are ``confined to a certain time window''. We think that this paper has the effect of suggesting the plausible conjecture that actually this time window is of infinite size. Let us see why. We notice that in a sense the memory kernel of Eq.\ (\ref{result}) affords a way to measure the aging process. In fact, if the waiting time distribution $\psi(t)$ is exponential, the memory kernel is equivalent to that produced by a bath that would regress to equilibrium instantanesously, with no aging effect. The aging effect, from within our perspective, can also be interpreted as a regression to equilibrium of a system of interest, under the influence  of a bath that is  regressing, too, to equilibrium. In the Poisson case, the bath regression to equilibrium would be instantaneous, thereby yielding no aging. In the case of $\mu < \infty$ we have a memory lasting forever. This is evident when $\mu < 3$, since in this case the adoption of Eq.\ (\ref{result}) proves that the Laplace transform of the memory kernel,  $\hat \Phi (u)$, tends to vanish for $u \rightarrow 0$.  This is an indication of infinitely persistent oscillations that are clearly incompatible with the Markov approximation.
\\ 
    
  Apparently, the condition $\mu > 3$ seems to be compatible with the Markov approximation, given the fact that in that case, $\hat \Phi (u)$ tends to a finite value for $u$ tending to $0$. We note that, while the analytical expression of $\Phi(t)$ is at the moment not yet known to us, the analytical form of $\Phi_{\xi}(t)$ is under our total control, and it is a an inverse power law with index $\beta = \mu -2$. The free diffusion process resulting from the fluctuations of $\xi(t)$ is known. 
The pioneering work of Ref.\ \cite{klafter} proves that even in the case $\mu > 3$ the Gaussian condition is not exactly realized. The central part of the diffusion pdf is the ordinary Gaussian diffusion, but at large distances slow tails with an inverse power law appear. These tails  become weaker and weaker upon increase of $\mu$ and might be annihilated by even weak external fluctuations, with the ensuing annihilation of aging.
 However, there are deep physical motivations to fully understand the theoretical issues generated by the aging process. A relevant example of this theoretical request is given by the physics of glassy systems, which are characterized by aging  and consequently by a violation of the ordinary fluctuation-dissipation condition, which is intimately related to the validity of the Onsager principle \cite{crisanti}.  In other words, we have to do further research work to understand how robust the non-Poissonian nature of $\psi(t)$ is, and to establish the intensity of environemental  fluctuations that are expected  to produce a truncation of the inverse power law nature of $\psi(t)$. This can be done along the lines of Ref.\ \cite{elena}. This research work can be carried at a much deeper level, as a consequence of the fundamental result of Eq.\ (\ref{result}). In fact, this equation defines the memory kernel $\Phi(t)$ of the GME, in terms of the waiting function $\psi(t)$.  We assume that the function $\psi(t)$ can be derived from the experimental observation  as in the case of the blinking quantum dots of Ref.\ \cite{chemphys}. Thus, with the help of Eq.\ (\ref{result}) we can draw from the experimental function $\psi(t)$ information on the bath responsible for the intermittent character of the process under study. This information will help the foundation of a dynamic model, and consequently will make possible to discuss the robustness against noise of the intermittent process under study, from a perspective more realistic than in the earlier work, for instance, the work of Ref.\ \cite{elena}.

Let us summarize the main results of this paper. The functions $\Phi(t)$ and $\psi(t)$ are the main mathematical properties behind the GME and CTRW, respectively, the former affording indications on the complex dynamics of the irrelevant variables and the latter illustrating the nature of the intermittent process that can be observed as a result of a special experimental detection \cite{chemphys}. The deviation of $\psi(t)$ from the exponential condition is equivalent to forcing $\Phi(t)$ depart dramatically from the white noise condition. The non-Poisson nature of $\psi(t)$ generates memory infinitely extended in time, and aging with no finite time scale as well, no matter how big the scaling parameter $\mu$, provided that $\mu < \infty$. 
 In fact, if a finite time scale for both aging and memory existed, at a much larger time scale, the function $\Phi(t)$  would be equivalent to a delta of Dirac, the bath fluctuations would be white noise, the bath would be at equilibrium, and, of course, $\psi(t)$ would be exponential.  This is equivalent to proving that the non-exponential nature of $\psi(t)$ yields aging effects that are not limited to the non-stationary case $\mu < 2$, even if in this case these aging effects seem to be more natural. The condition $\mu > 3$ is compatible with the stationary condition. However, a bath generating this condition, through the key result of Eq.\ (\ref{result}), is characterized by dynamic properties with memory infinitely extended in time.

\begin{acknowledgments}
We thank Professor Bill Schieve for his remark that we are using a generalized form of Onsager's
principle. We thankfully acknowledge the financial support of ARO, through Grant DAAD19-2-0037.
\end{acknowledgments}

\end{document}